\documentclass[a4paper,fleqn,usenatbib]{mnras}
\usepackage[dvipsnames,usenames]{color}
\usepackage{mathptmx}
\usepackage{etoolbox}
\usepackage{graphicx}
\usepackage{amsmath}
\usepackage{amssymb}	

\newcommand{\rd}{{\rm d}}
\newcommand{\vect}[1]{{\mathbf{#1}}}

\usepackage{amsmath}	
\pubyear{2019}

\title{A Markov Chain Monte Carlo approach for measurement of jet precession in radio-loud active galactic nuclei}

\author[M. A. Horton et al.]{
Maya A. Horton,$^{1}$\thanks{E-mail: mh17adw@herts.ac.uk}
Martin J. Hardcastle,$^{1}$
Shaun C. Read$^{2}$
and Martin G. H. Krause$^{1}$
\\
$^{1}$Centre for Astrophysics Research, School of Physics, Astronomy
and Mathematics, University of Hertfordshire, College Lane, Hatfield, AL10 9AB, UK\\
$^{2}$INAF -- Osservatorio Astronomico di Roma, via Frascati 33, I-00040 Monte Porzio Catone, Roma, Italy \\
}

\date{Accepted XXX. Received YYY; in original form ZZZ}

\pubyear{2019}
\hypersetup{draft}

\begin{document}
\label{firstpage}
\pagerange{\pageref{firstpage}--\pageref{lastpage}}
\maketitle

\begin{abstract}
Jet precession can reveal the presence of binary systems of
supermassive black holes. The ability to accurately measure the
parameters of jet precession from radio-loud AGN is important for constraining the binary supermassive black hole population, which are expected as a result of hierarchical galaxy evolution. The age, morphology, and orientation along the line of sight of a given source often result in uncertainties regarding jet path. This paper presents a new approach for efficient determination of precession parameters using a 2D MCMC curve-fitting algorithm which provides us a full posterior probability distribution on the fitted parameters. Applying the method to Cygnus A, we find evidence for previous suggestions that the source is precessing. Interpreted in the context of binary black holes leads to a constraint of parsec scale and likely sub-parsec orbital separation for the putative supermassive binary.
\end{abstract}


\section{Introduction}
\label{sec:intro}
The detection of gravitational waves from black hole mergers has brought new opportunities for exploring the coalescence of binary systems \citep{abbott16b, abbott16a, abbott17}. However, current instrumentation is limited to the detection of stellar-mass black hole mergers. Given that mergers between galaxies are common and play a fundamental role in galactic evolution \citep[e.g.,][]{rodriguez15,carpineti15}, the case has been made that many galaxies may contain binary supermassive black holes \citep{begelman80,mayer17,tremmel18}. 

If a binary black hole produces a jet, the jet will exhibit long-term precession due
to the geodetic precession mechanism \citep{begelman80}. Jet curvature
on large physical scales
can therefore in principle be used as a mechanism for determining
precession period, and hence the angular separation between objects in
a binary (see equation~\ref{eq:krause_eq2}, reproduced from \citet[hereafter K19]{krause18}). This can be achieved by identifying observable signatures of
precession, which appear with different morphological characteristics.

Such observational markers include S-shaped symmetry between jet and
counterjet; curvature of jet (which can vary from straight to highly
curved); multiple or ring-shaped hotspots, and misalignment between
jets and lobes, such as the occurrence of the jet towards the edge of
the lobe. A recent
analysis of nearby ($z < 1$) radio sources found precession markers in
73 per cent of examined sources (K19). The consequences of jet precession are far-reaching: AGN are partly responsible for heating the intra-cluster medium \citep[ICM,][]{croton06, best07, raouf17, turner15}, but radio jets may not adequately heat the ICM \citep{hardcastle13} without an additional mechanism, such as varying jet directions \citep{babul13}. Accurate predictions of jet morphology may allow for better modelling of AGN feedback effects.

In the coming decades, a new generation of gravitational wave
detectors will allow us to detect both transient and continuous
sources of gravitational waves \citep{amaro12}. Even the most massive
supermassive black holes will be detectable via the Square Kilometer
Array's pulsar timing array \citep{wang17,stappers18}. We can hope to
gain first insights into the binary supermassive black hole population
from an analysis of the morphology of extragalactic radio sources.
Data from surveys such as LOFAR \citep[e.g.,][]{shimwell19} have
vastly increased the number of known radio-loud AGN sources in the
nearby universe, many of which contain complex morphologies which may
exhibit potential precession indicators \citep[e.g.,][]{hardcastle19}.
High resolution VLA data indicates potential complexities in jet
structure, some of which may be candidates for jet precession \citep[e.g.,][]{mahatma19}.

Doppler boosting and relativistic aberration influence the observed jet structure. This becomes more pronounced at higher inclination angles, to the point where the counterjet may be rendered invisible. Conversely, hydrodynamic interactions between jet, lobe, and ICM can result in light jets being pushed in one direction or another. Here, we will assume that hydrodynamic interactions are negligible, which will be discussed later.

 The aim of this paper is to provide an efficient, novel approach to constrain the precession parameters in radio jets which have already been identified as precession candidates. Assuming ballistic precession of a jet emanating in the spin direction from a spinning member of a supermassive binary that undergoes geodetic precession, the precession parameters can be turned into constraints on the binary orbit, thus informing interpretation of future gravitational wave observations. We will present its efficacy on noisy data, as well as for sources where knowledge of the jet path is incomplete, and trial it on Cygnus A.

\section{Simulations}
\label{sec:mcmc}
\subsection{MCMC Approach}
In this project we use a Markov-Chain Monte Carlo (MCMC) approach for jet path determination, based on the Python \verb emcee  package \citep{foremanmackey13}. This makes use of the Goodman and Weare affine-invariant ensemble sampler \citep{goodman10}, which is valuable for marginalising over nuisance parameters from complex models with high-dimensional data, as is the case here. 

For nearby ($z < 1$) sources with well-known priors (such as Cygnus A)
and a high number of easily-identifiable jet knots (typically where
the number of observed jet points is higher than the number of model
parameters, including nuisance parameters), well-fitting jets can be
found by minimising $\chi^2$ in a manual or brute-force search of
  parameter space. However, since the observed sources have complex
geometry, and many have not been studied enough to have good
constraints on priors, the approach rapidly becomes ineffective in
finding the best fits, in the sense that it is hard to demonstrate
that a true global minimum has been found, and also
  computationally infeasible in the case where broad priors require a
  large parameter space to be searched.

Our model instead works on the assumption that the likelihood
$L$ is proportional to the line integral through probability density
space:
\begin{equation}
L(p_1,p_2\dots) \propto \int_M p(\vect{x}) \rd x 
\end{equation}
where $\vect{x}$ is a position in two-dimensional space, $\rd x$ is
the scalar line element, the integral is evaluated numerically over the jet path
$M$ defined by the model parameters $p_1, p_2\dots$ (see the following
subsection) and $p(\vect{x})$ is the sum of the probability densities
due to all data points at $\vect{x}$. That is, if there are $n$ jet knots at positions $\vect{d}_1, \vect{d}_2\dots$ and the error on all
positional measurements is taken to be $\sigma$, then
\begin{equation}
p(\vect{x}) = \frac{1}{n} \sum_{i=1}^{n}
\exp\left(-\frac{|\vect{x}-\vect{d}_i|^2}{2\sigma^2}\right)
\label{eq:likelihood}
\end{equation}
In the numerical evaluation of the integral we break the path up into
finitely many points and then iteratively increase the sampling until
convergence is reached to a given tolerance level, typically for around 1000 points along the path.

An important detail is the use of a prior inversely proportional to
the length of the jet (i.e. $\int_M \rd x$). This ensures that the
code will not favour arbitrarily long jets that maximize $L$ by
passing close to each data point many times. We also included one
extra variance parameter $V$, with a half-Cauchy prior, which is added in quadrature to the
measurement error on the data points $\sigma_d$ as discussed by e.g.
\cite{hogg10}: that is, $\sigma^2 =
\sigma_d^2 + V$. This improves the
convergence of the algorithm in the burn-in phase by preventing it from
becoming stuck in local minima. This parameter always converges on a
very low value by the time burn-in is complete and so we treat it as a
nuisance parameter that can be marginalized over in the remainder of the paper.

\subsection{Precessing Jet Model} \label{subsec:model}

\citet{gower82} developed a relativistic jet curvature model for precessing jets where light travel time is close to jet expansion time (resulting in relativistic aberration effects). They give the instantaneous velocity vector as follows: 

\begin{align}
v_{\mathrm{x}} &= s_{\mathrm{jet}}\beta c \left\{\sin~\psi~\mathrm{sin}~i \, \mathrm{cos} \left[\Omega (t_{\mathrm{ej}}-t_{\mathrm{ref}})\right] + \mathrm{cos}~\psi~\mathrm{cos}~i \right\} \\
v_{\mathrm{y}} &= s_{\mathrm{jet}} \beta c~\mathrm{sin}~\psi~\mathrm{sin}~\left[\Omega(t_{\mathrm{ej}} - t_{\mathrm{ref}})\right] \\
v_{\mathrm{z}} &= s_{\mathrm{jet}} \beta c \left\{ \mathrm{cos}~\psi~\mathrm{sin}~i - \mathrm{sin}~\psi~\mathrm{cos}~i~\mathrm{cos}~\left[\Omega(t_{\mathrm{ej}} - t_{\mathrm{ref}}) \right] \right\}
\end{align}
where $s_{\mathrm{jet}}$ is a sign parameter corresponding to 1 for the jet, and $-1$ for the counterjet; $\beta = v/c$ where $v$ is jet speed; $i$ corresponds to the inclination angle of the jet along the line of sight; $\psi$ corresponds to precession cone opening angle; $t_{\mathrm{ej}}$ is the ejection time of an individual plasmon (packet of radio energy), and $t_{\mathrm{ref}}$ is a reference time taken some time after $t_{\mathrm{ej}}$. $\Omega$ is the precession frequency defined as $\Omega = 2\pi/p$ (where $p$ is precession period), and is given in units of radians per second. 

From here, the jet path on the sky can be found: 
\begin{align}
\phi_{\mathrm{z}} &= v_{\mathrm{z}}(t_1 - t_{\mathrm{ej}})/\left[d(1 - v_{\mathrm{x}}/c) \right] \\
\phi_{\mathrm{y}} &= v_{\mathrm{y}}(t_1 - t_{\mathrm{ej}})/\left[d(1 - v_{\mathrm{x}}/c) \right]
\end{align}
where $\phi_{\mathrm{z}}$ and $\phi_{\mathrm{y}}$ correspond to jet
angular motion as viewed by the observer: in the notation of the
previous Section,
$\mathbf{x} = (\phi_{\mathrm{y}}, \phi_{\mathrm{z}} )$, up to some unknown position angle $\alpha$ on the
sky. The jet path is obtained by fixing $t_1$, the epoch of observation, and varying $t_{\mathrm{ej}}$ from $t_1$ to $0$, to build up a picture of the jet. Phase angle, $\theta$, at the black hole at the time of observation, is given by $\Omega( t_1 - t_\mathrm{ref})$.
    
\begin{figure}
      \centering
 	\includegraphics[width=0.4\textwidth]{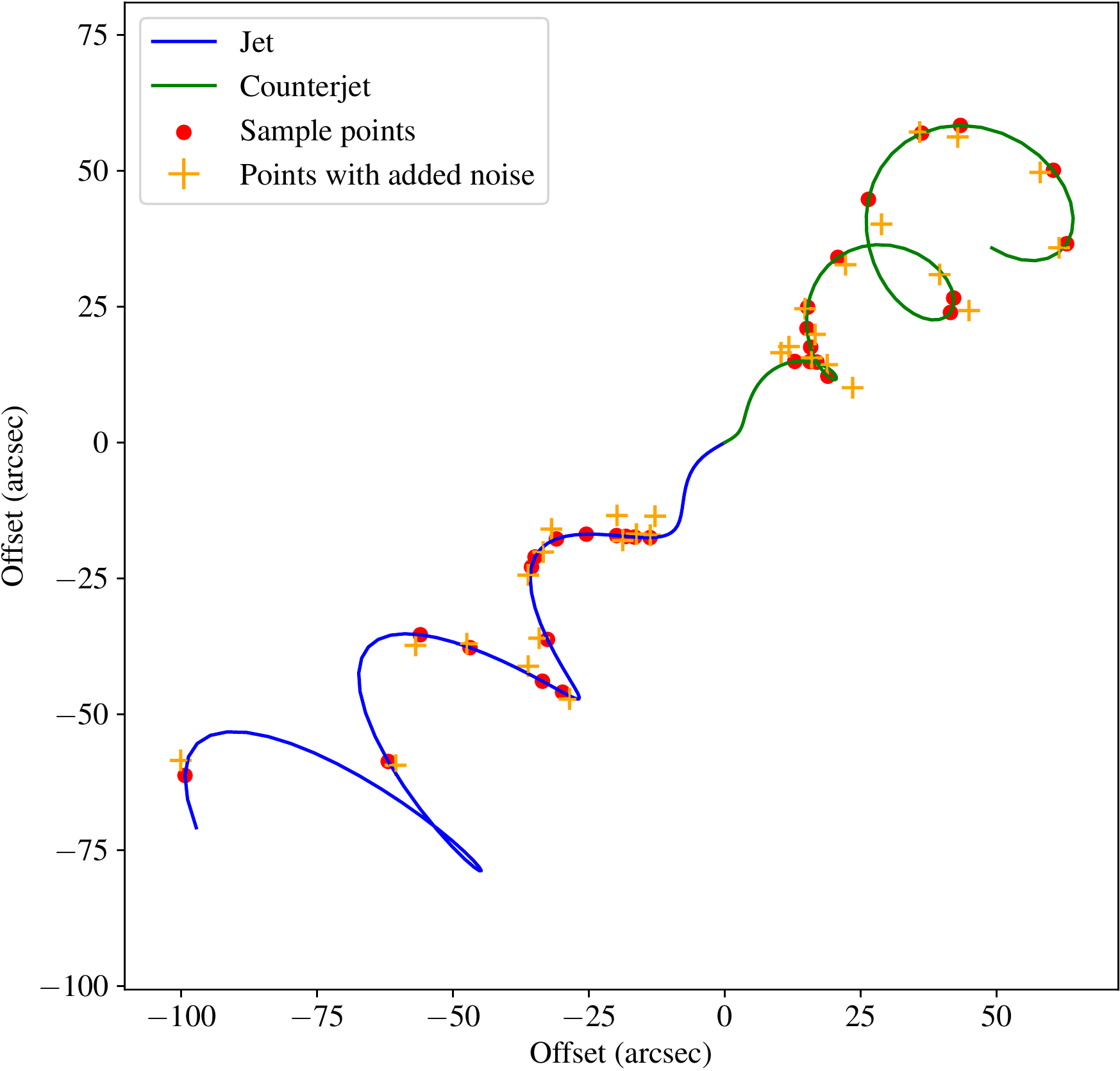}
     \caption{Generated data points for a simulated jet and
       counterjet projected on the sky. This particular example has
       $i=70^\circ$, $\psi=15^\circ$, $\theta = \pi/2$ radians,
       log precession period of $-0.5$, jet speed of $0.6c$, and
       $\alpha = 135^\circ$ radians. We assume a jet lifetime of 1 Myr
       and the source is taken to be at $z=0.1$. This example uses 15 data points for
       each side, resulting in 30 jet points ($d_i$ from equation~\ref{eq:likelihood}). Red circles show randomly
       generated points drawn from the jet/counterjet path shown in
       blue; orange crosses show the effect of adding independent
       Gaussian noise to the $x$ and $y$ positions, here with $\sigma
       = 2.0$ arcsec.}
     \label{fig:cjet_generate}
\end{figure}

A simulated jet path is shown in
Fig.~\ref{fig:cjet_generate}. This shows the path for a jet with $t_1
= 1$ Myr, $\log_{10}(p/\rm{Myr}) = -0.5$, $\beta=0.6$ and $z=0.1$.

\section{Results}
\subsection{Model Validation} \label{subsec:param_study}

Model validation was carried out using simulated data. We simulated
jets using the model of Section \ref{subsec:model}. We drew data
points (positions on the jet, the $\vect{d}_i$ of
Section\ \ref{sec:mcmc}) at random from the length of the simulated
jets in order to 
represent the fact that real jets are knotty and not detected
continuously along their length. We added
independent Gaussian noise in the $x$ and $y$ directions on the sky to
the $\vect{d}_i$ to
mimic the effect of observational uncertainties on the data.
Fig.~\ref{fig:cjet_generate} shows the relationship between simulated
data and the original samples from the jet and counterjet. In our
validation tests we used a model very similar to that plotted in
Fig.\ \ref{fig:cjet_generate}: in particular all test simulations had
$\beta=0.6$, $t_1 = 1$ Myr and $z=0.1$, as these are reasonable
parameters for the type of jet we hope to study. The phase angle
$\theta$ was fixed to $\pi/2$ radians as this simply corresponds to a
rotation of the precessing jet about its axis and should not affect
our ability to recover other jet parameters by fitting.

Initially we focused on simulated sources where only a single jet is
  detected. We produced three groups of simulated precessing jets, each
with 30 fitted jet points. For each set, we varied one parameter and
kept the others constant. Unless otherwise specified we use an
inclination angle of 70$^\circ$, an opening angle of 15$^\circ$ and a
precession period $p$ such that $\log_{10}(p)=-0.5$, as shown in
Fig.\ \ref{fig:cjet_generate}. The MCMC fitting used 96 walkers and
ran for 5000 steps; the first 400 were removed as burn-in. Initial
  positions of the walkers are drawn uniformly from the priors on each
  parameter except for jet length (as described in Section \ref{sec:mcmc}. We verified by inspection of the tracks taken by the 
walkers that 400 steps was a conservative value to use for burn-in.
This search used flat priors.

\begin{table}
\caption{List of parameters varied during parameter space search. For each combination of parameters, 40 realizations of random point generation and MCMC fitting were produced.}
\centering
\begin{tabular}{l c c}
\hline\hline
Parameter & Values & Units \\ [0.5ex] 
\hline
Inclination angle & 30, 60, 90 & Degrees \\
Precession cone opening angle & 15, 30, 45 & Degrees \\
Precession period & -0.5, 0, 0.5 & $\rm log_{10}({\rm p/Myr})$\\
\hline
\end{tabular}
\label{table:param}
\end{table}

\begin{figure}
 	\includegraphics[width=\columnwidth]{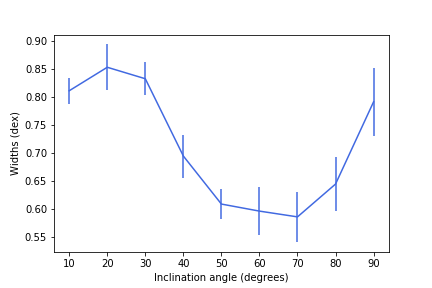}

 	\includegraphics[width=\columnwidth]{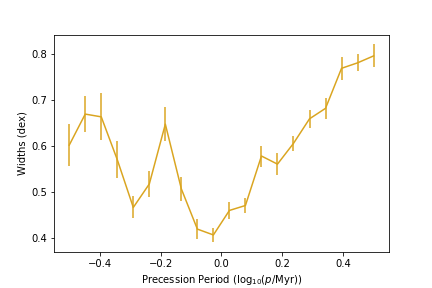}

 	\includegraphics[width=\columnwidth]{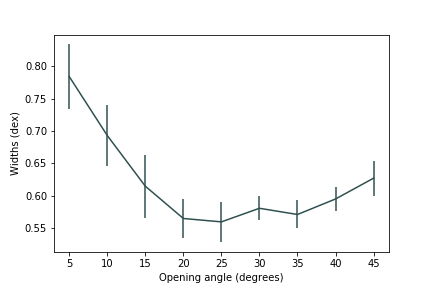}
     \caption{Results of parameter study. Each plot shows the mean width of
       credible intervals for the precession period in dex -- i.e.
       twice the $1\sigma$ uncertainty in dex -- over each
       parameter that was varied in the specific series of runs. Top
       panel: inclination angle varies from $10^\circ$ to $90^\circ$.
       Middle panel: precession period varies within $-0.5 < \log (p /
       \mathrm{Myr}) <0.5 \log$. Bottom panel: precession cone opening
       angle ranging from $5^\circ$ to $45^\circ$. The constraints on
       precession period are good (uncertainties less than a factor
       $\sim 3$) across all of the parameter space,
       with smaller credible intervals being found at higher inclination angles and with a precession period closer to 1 Myr. } 
     \label{fig:truth_all}
\end{figure}

For each combination of input model parameters as listed in Table~\ref{table:param}, we generated 40 instances of
  simulated data (as shown in Fig.\ \ref{fig:cjet_generate}), where each instance generates a different combination of randomly-generated points and noise. We ran
  the MCMC fitting on each simulated dataset, fitting for the
  inclination angle $i$, the cone angle $\psi$, the phase $\theta$,
  the precession period $\log(p)$, the jet speed $\beta$ and the
  position angle on the sky $\alpha$. To characterize the
quality of the fits we took the widths of the distribution functions
of the precession period, defined as the distance between the upper
and lower bounds of the credible intervals. The credible interval is
defined as the 68 per cent confidence interval around the peak of the
posterior distribution (the highest posterior density interval).
As our figure of merit we used the credible interval on the posterior probability
  distribution for precession period, which was invariably peaked
  close to the true value. For a given set of model parameters, we
  recorded the means of the widths of the credible intervals on
  precession period.
Standard deviations of the measured widths are roughly 10 per cent of
the widths or less. For inclination angle, we varied the inclination
from 10$^\circ$ to 90$^\circ$. For precession period, we varied the
logarithm of the precession period in Myr from $-0.5$ to 0.5. Finally,
for the cone opening angle, we varied the precession cone opening
angle from 5$^\circ$ to 45$^\circ$. Fig.~\ref{fig:truth_all} shows our
results. The mean credible interval for our simulated data varies
between 0.4 and 0.8, corresponding to uncertainty factors of 1.6 to
2.5. Smaller credible intervals (i.e. better-constrained precession
periods) are obtained with inclination angles from 30$^\circ$ to
80$^\circ$, precession periods in the range 0.5 Myr to 1 Myr, and
opening cone angles greater than 15$^\circ$.

We conducted a full quantitative search of the parameter space and
found it possible to find a peaked posterior distribution for
precession period using this method, alongside good constraints on
other parameters. We get particularly good constraints (i.e. narrow
posterior probability distributions centred round the true values) on
position angle $\alpha$, phase $\theta$, and precession cone opening
angle $\psi$. The parameter of greatest interest is precession period,
as this is the one from which we can obtain information regarding
binary separation. We verified that for these simulations we obtain
unbiased estimates of the precession period using the median of the
posterior probability distribution, as expected.

\begin{figure}
 	\includegraphics[width=\columnwidth]{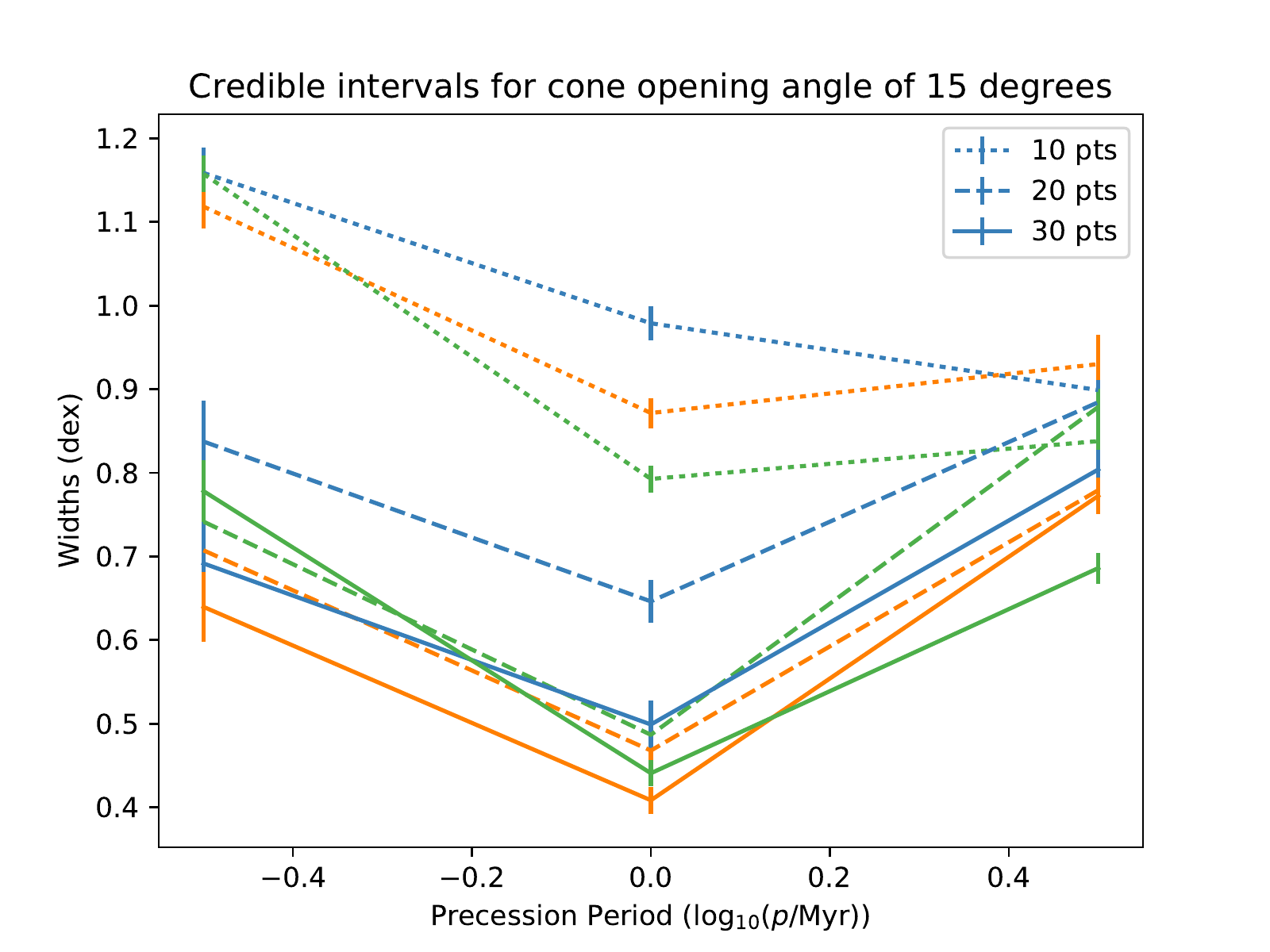}

 	\includegraphics[width=\columnwidth]{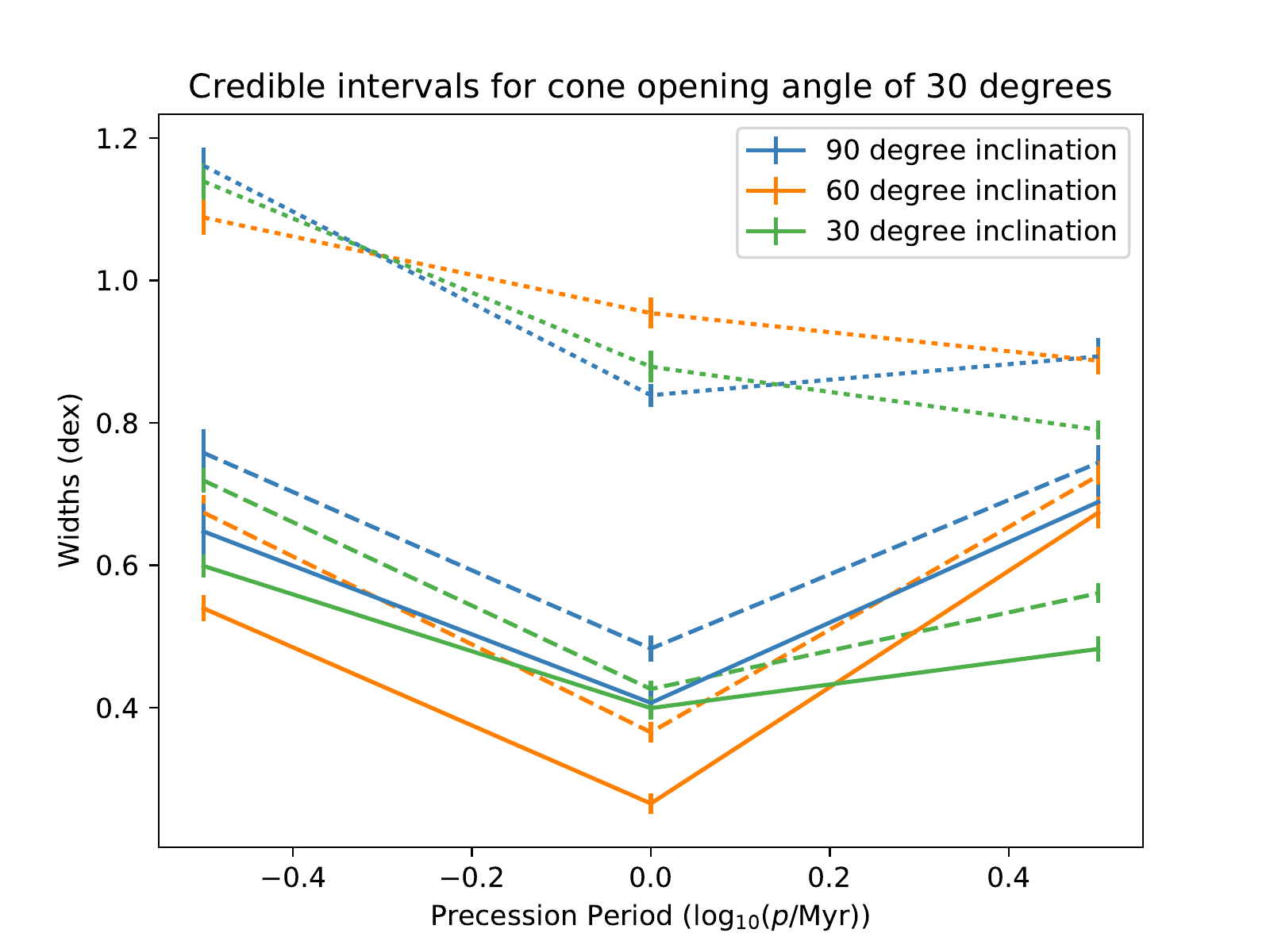}

 	\includegraphics[width=\columnwidth]{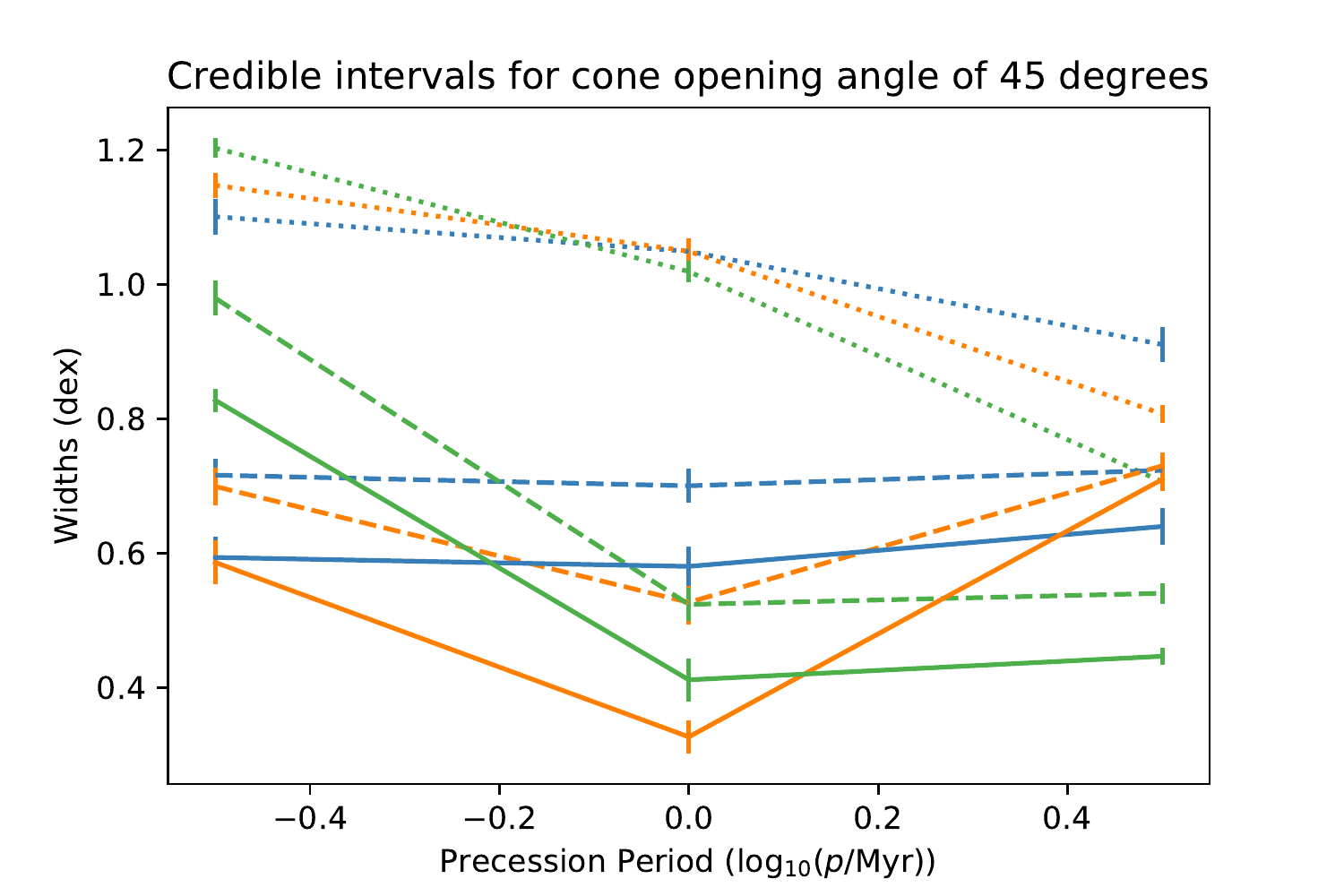}

     \caption{Investigation of the number of jet points required to obtain a certain fit quality depending on input parameters. From top to bottom, the precession cone opening angle changes from 15$^\circ$ over 30$^\circ$through 45$^\circ$. Blue lines correspond to inclination angles of 90$^\circ$, orange are at 60$^\circ$ and and green at 30$^\circ$. Solid lines represent a fit of 30 jet points, dashed lines are 20 and dotted are 10. 20 data points show a marked difference in credible interval widths compared to ten points, whilst the addition of another 10 has less of an impact.}
     \label{fig:incl_comparisons}
\end{figure}

We investigated how the number of fitted jet points affects fit
quality. We therefore repeated the study, but for 20 and 10 simulated
points. The result is shown in Fig.~\ref{fig:incl_comparisons}. We
found that 20 jet points performed similarly to 30 points. However,
the quality of fits, in terms of the constraints that we obtained on
precession period, decreased markedly with 10 points. This is not
surprising given that the number of jet points becomes close to the
number of degrees of freedom of the model (6).

\subsection{Counterjet}\label{subsec:cjet}

\begin{figure}
 	\includegraphics[width=\columnwidth]{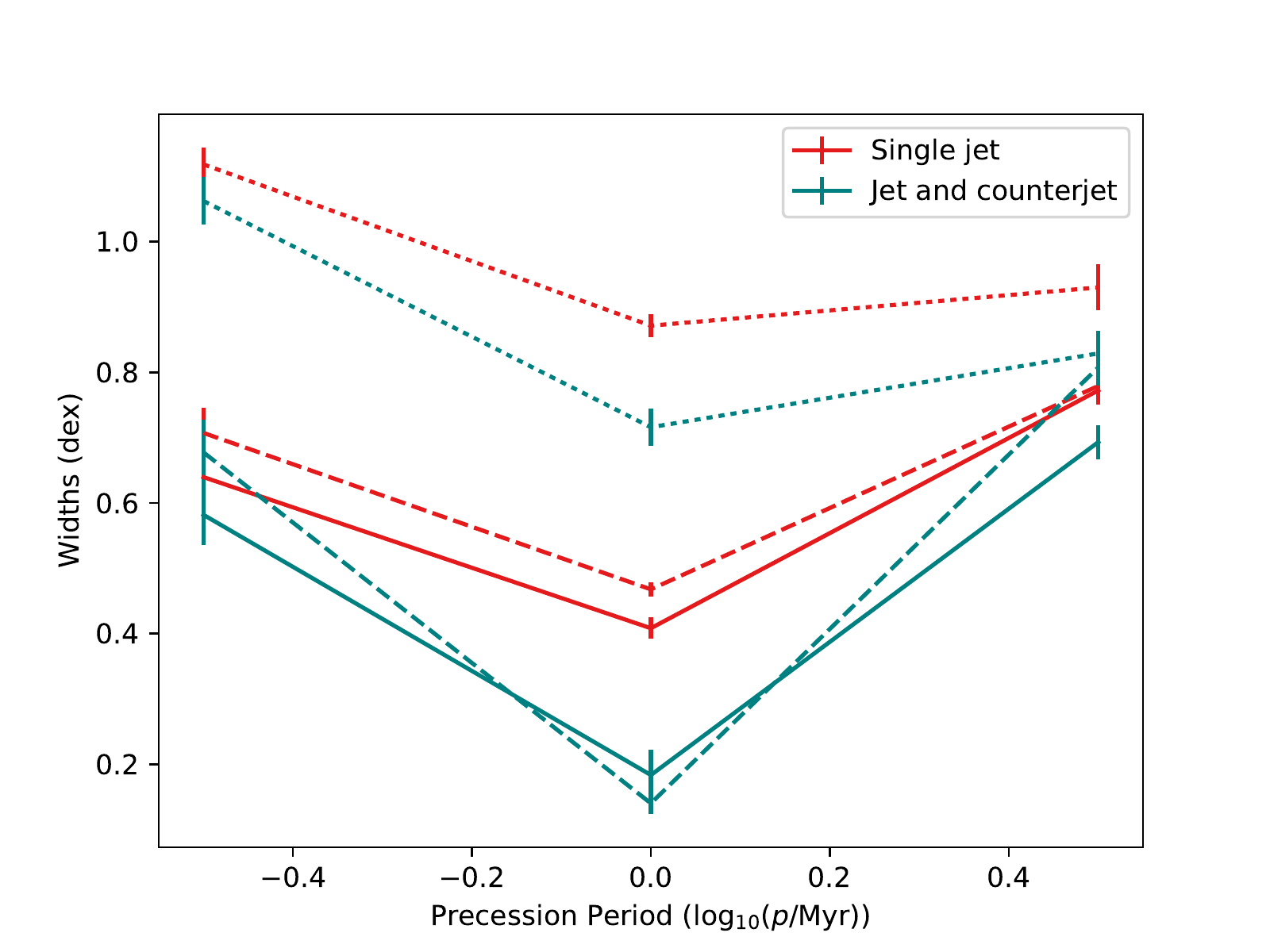}
     \caption{Comparison showing credible intervals counterjet (teal) and single jet (red) for an inclination angle of 60 degrees and a precession cone angle of 15 degrees. Solid lines represent a fit of 30 jet points, dashed lines are 20 and dotted are 10.}
     \label{fig:jet_comp}
\end{figure}

 All modelling up to here was done with a single jet, the
   approaching jet: in many real sources the approaching jet is the
   only visible one because of the strong effects of Doppler boosting. We also ran
models where we distributed the same number of jet points 
either on the approaching jet, only, or on both, jet and counterjet
in order to find out whether having the same number of points spread over two jets gave a better, worse, or equal fit. Fig.~\ref{fig:jet_comp} compares a single jet model (red) and one containing both jet and counterjet (teal; see Fig.~\ref{fig:cjet_generate} for an example of simulated data which includes the counterjet.).

For the same set of parameters (here, using one of the better fits
identified in the previous parameter study, of $i=60^\circ$ and $\phi
= 15^\circ$), we observe that the inclusion of a counterjet in an
otherwise-identical model produces consistently better constrained
precession period estimates, by up to
$0.3$ dex, across 30, 20, and 10 points. Again, fits with 10 data
points give only poor constraints on precession period, but the
presence of a counterjet still results in an improvement of $0.1$ dex.

Given the relativistic nature of the jet path, we suggest that
  the improvement when a counterjet is included comes
from the additional morphological constraints provided by the receding
jet, which is not simply an inverted copy of the jet (as seen in
Fig.\ \ref{fig:cjet_generate}).

\section{Cygnus A}
We use the well-studied extragalactic radio source Cygnus~A to demonstrate the applicability of our code. This is one of the nearest FRII galaxies and has a well-defined jet and counterjet with a jet inclination angle thought to be between 55-85$^\circ$ \citep{bartel95}, with a redshift of $z=0.0562 \pm 0.000067$ \citep{carilli96b}. There is observational morphological evidence to suggest that the jets are precessing (K19). 

We visually identified 48 jet points (with an astrometric error of 0.4
arcsec from VLA data) from the 5-GHz map of \cite{perley84}. Jet
points are taken to be any discrete, distinct feature that appears to
be part of the jet; this includes both compact knots and sub-regions
of more extended bright parts of the jet. For longer sub-regions of
the jet we placed multiple points, spaced by at least the resolution
of the image, in order to constrain the curvature of the jet in these regions.

\begin{figure*}
 	\includegraphics[width=\linewidth]{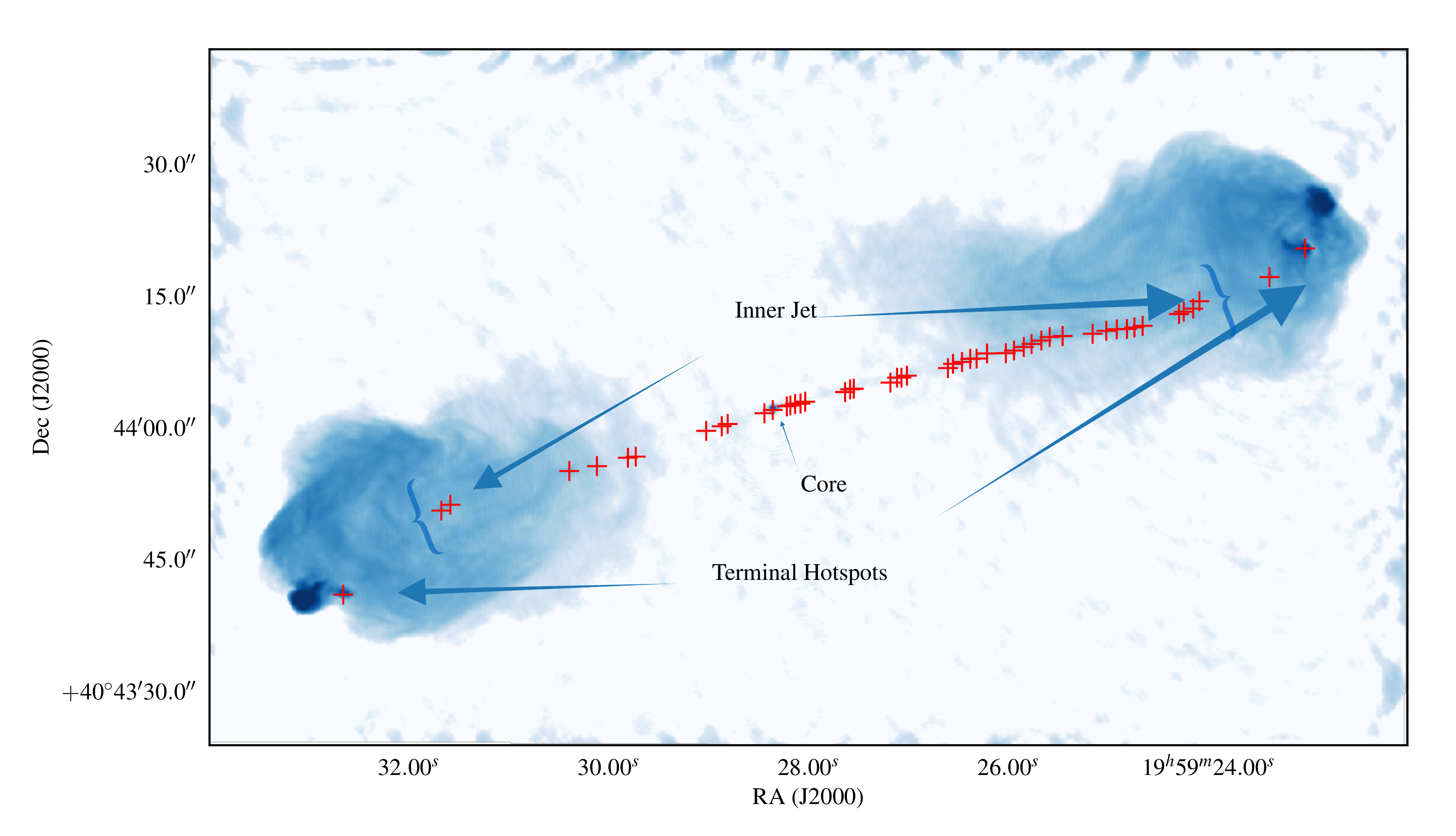}
    \caption{Locations of visible jet knots within Cygnus~A used in
      the model. Note the curvature in both the jet and counterjet
      paths. The final points in both jet and counterjet lie
        above the primary hotspots of the two lobes as discussed in
        the text. The inner, straighter points of the jet used in Fig.~\ref{fig:StraightJet} are identified as being within the indicated brackets.}
    \label{fig:cygA_points}
\end{figure*}

One important constraint on the application of the MCMC model to
real-world data is the requirement for the precessing jet to stay
within the lobes. Since the model doesn't take lobe structure into
account beyond the terminal hotspot, the constraint was implemented by
a joint prior on position angle and precession cone opening angle.
This dual parameter calculates the two extreme edges of the cone for
any given position angle, and keeps them within the lobes, which are
approximately $25^\circ$ across at their widest point. Without
  this additional prior, we found that unphysical jet paths were permitted as
  part of the fit, broadening the credible interval on precession period. As with the
tests described in earlier sections, the priors were flat within the
permitted ranges.

\begin{figure*}
 	\includegraphics[width=1.0\linewidth]{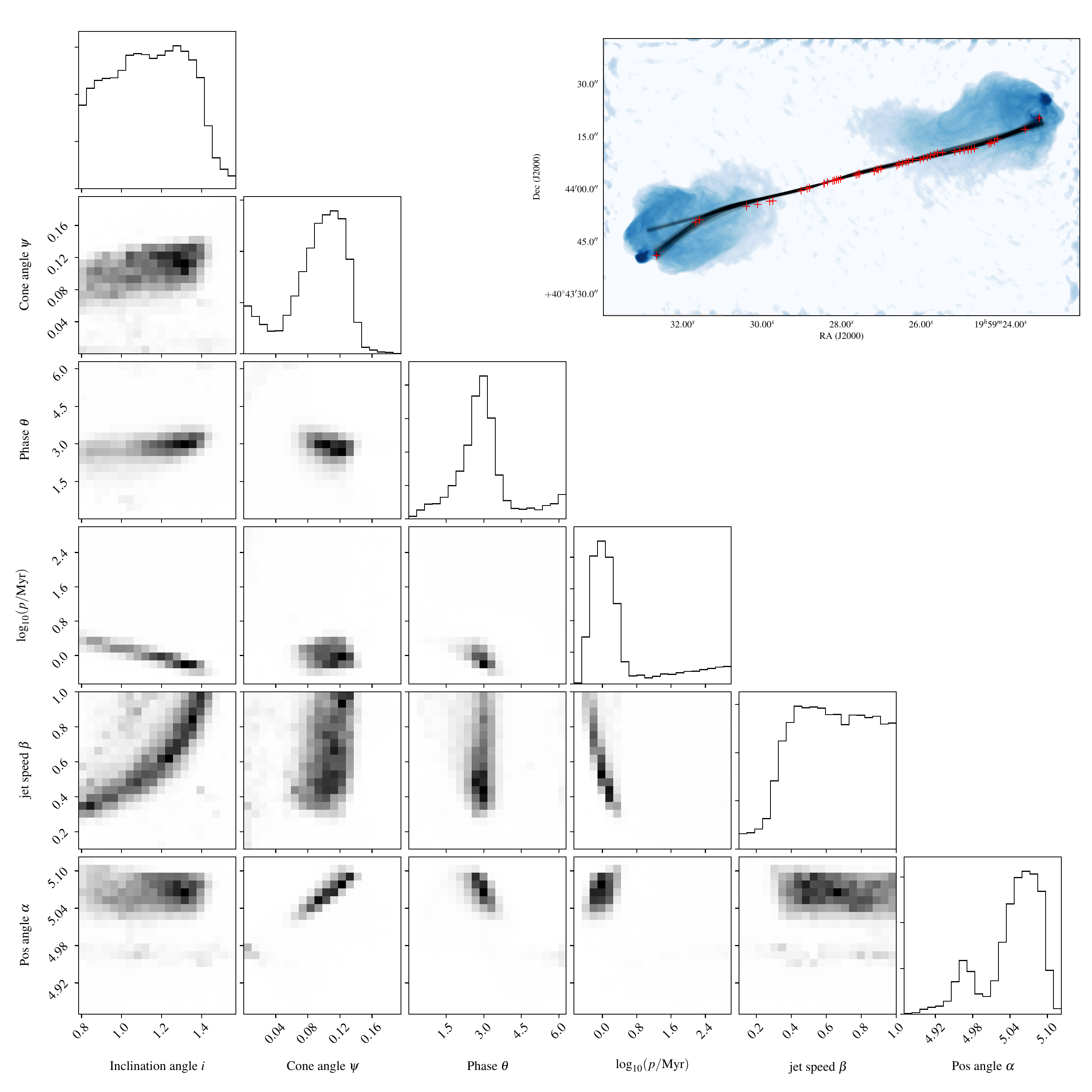}
    \caption{Main: `Corner plot' of one-dimensional and two-dimensional marginalized posterior probability distributions for the model parameters given the observed jet path of Cygnus~A, including all identified jet points up to terminal hotspot.  Inset: jet paths based on parameters drawn at random from the posterior overplotted on the radio image of Cygnus~A.}
    \label{fig:FullJet}
\end{figure*}

\begin{figure*}
 	\includegraphics[width=1.0\linewidth]{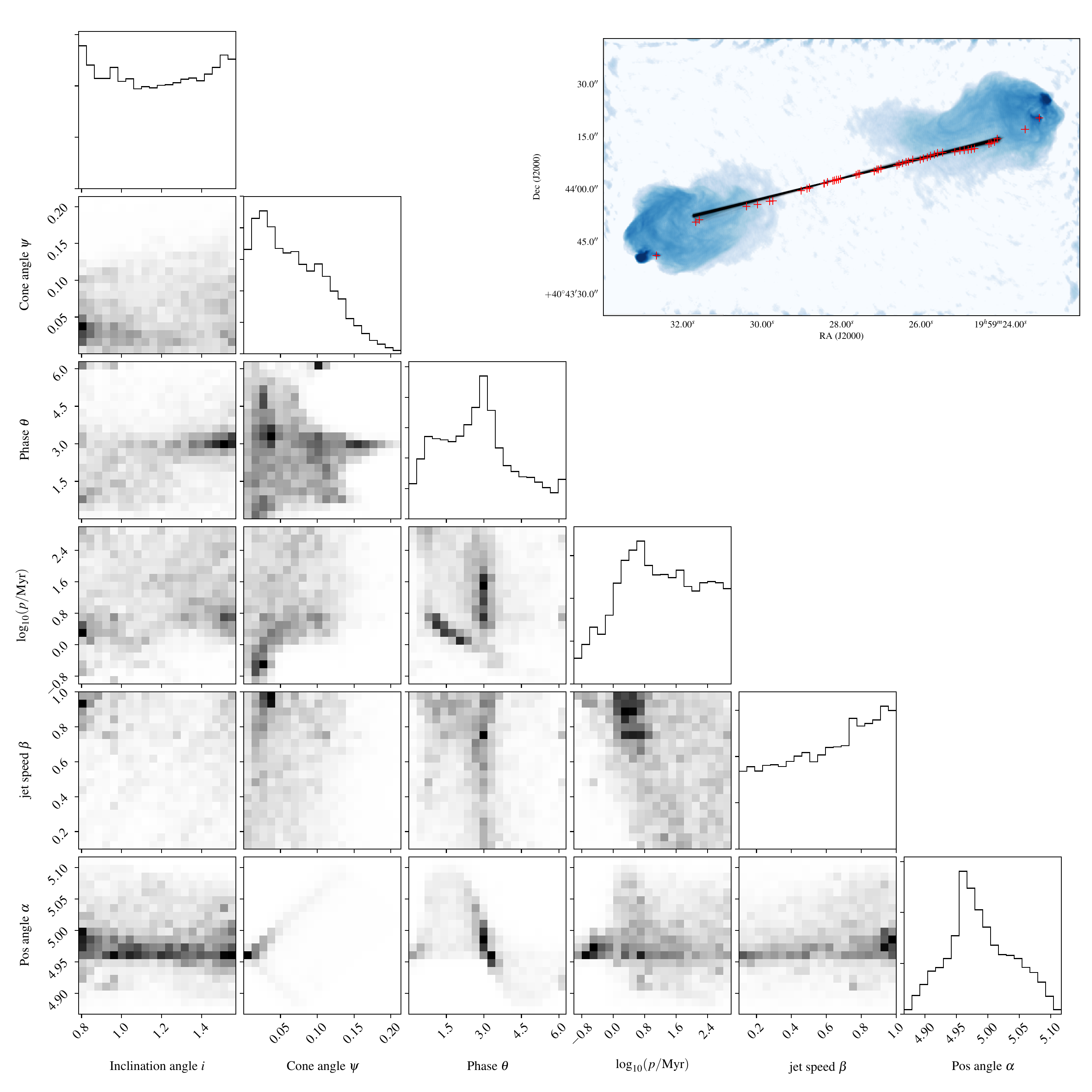}
    \caption{Main: `Corner plot' of one-dimensional and two-dimensional marginalized posterior probability distributions for the model parameters given the observed jet path of Cygnus~A, limited to straight portions of jet.  Inset: jet paths based on parameters drawn at random from the posterior overplotted on the radio image of Cygnus~A. The jet points used for this fitting are indicated by the extent of the overplotted jet paths.}
    \label{fig:StraightJet}
\end{figure*}

We initially fitted the model to all identified points (see
Fig.~\ref{fig:FullJet}), including the terminal hotspot, which in both
lobes is taken to be the more compact or `primary' hotspot; we do not
include the larger, brighter secondary hotspot since the position of
this is almost certainly dominated by post-shock hydrodynamics
\citep{cox91}. Then, with respect to the potential for hydrodynamic
influences to dominate the jet path within the lobes, fitted only to
the straight portion of the jet, terminating just inside the lobes
(Fig.~\ref{fig:StraightJet}). The full number of regions produced a
peak in the posterior distribution corresponding to a precession
period of 1~Myr. 

K19 show that, on the assumption that the jet is produced by the more
massive black hole, an upper limit on the binary separation in pc,
$d_{\rm pc}$, is given
by:
  \begin{equation}
    d_{\rm pc} < 0.18P^{2/5}_{gp,Myr}M_9^{3/5}
    \label{eq:krause_eq2}
\end{equation}
where $P_{gp,Myr}$ is the precession period in Myr and $M_9$ is the
black hole mass in units of $10^9 M_\odot$. Our estimated precession
period of $\sim 1$ Myr thus corresponds to an upper limit on the binary
separation distance of 0.3 parsecs. This was
calculated using the total central black hole mass of $(2.5 \pm 0.7)
\times 10^9 M_\odot$ \citep{tadhunter03}. Using only the straight part
of the jet within 50 arcseconds from the core results in a preferred range of 1-10~Myr for the precession period with less probability up to several 100~Myr. 

\begin{figure}
    \centering
    \includegraphics[width=\columnwidth]{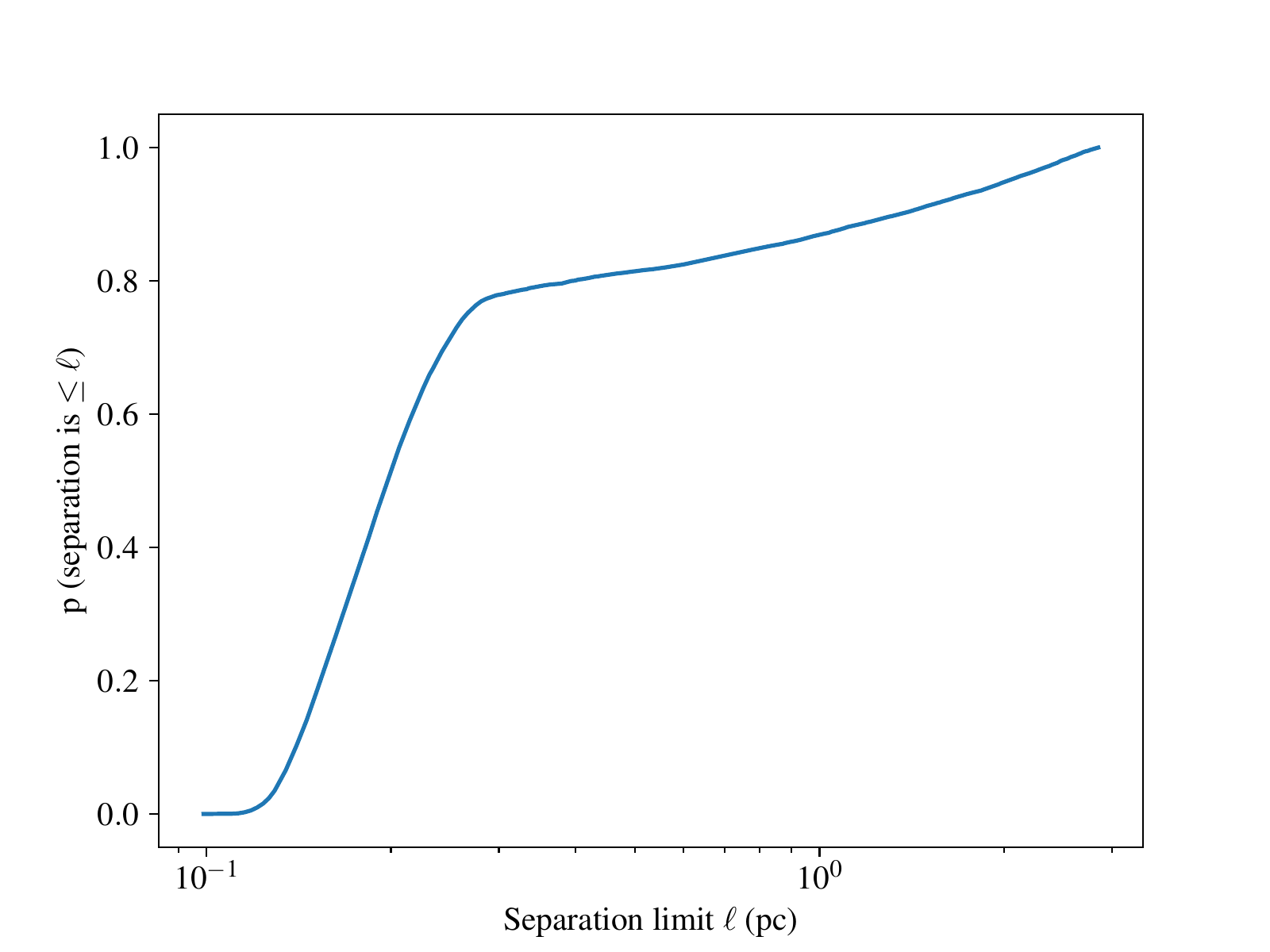}
    \caption{Cumulative probability of upper limit on binary separation calculated using equations in K19, using samples from the posterior with burn-in removed.}
    \label{fig:xlimit}
\end{figure}

Since the credible interval on the posterior encompasses a range of
precession period values, we can also consider the
constraints given by the full posterior rather than just its peak, as
seen in Fig.~\ref{fig:xlimit}. This shows the cumulative probability
distribution in log space of the binary separation limit as calculated from K19. This indicates that the lower limit of the cumulative probability of the separation being \textless 0.3 pc is 0.8, whilst \textless 1~pc is around 0.85, and at least 95\% confidence, the separation is \textless 2~pc. 


\section{Discussion}
We have shown that strong constraints on precession periods of
precessing jets can be obtained with MCMC fitting of the jet path, if
the jet is detected at at least 20 points. Even 10 points can yield
useful constraints. Furthermore, Fig.~\ref{fig:jet_comp} highlights that the presence of a counterjet produces much better fits in an otherwise identical model. That is, a single visible jet with 20 identifiable distinct regions produces less accurate constraints on the posterior distribution than 20 such regions distributed between the jet and counterjet. This is because of the different morphological distortion of jet and counterjet due to the relativistic aberration, which means that the counterjet is not just a mirror image of the jet. 

For 10 data points, we obtain a credible interval width of typically 1~dex, i.e.
an uncertainty of a factor 3 in either direction on the precession
period. For 20 data points, this is significantly improved, showing a
precision of around 0.4~dex. This is relevant for future binary black
hole research. Equation~\ref{eq:krause_eq2} shows that the separation of a binary
black hole can be constrained by the precession period, therefore
being able to adequately measure a precession period with
uncertainties at the level of 1~Myr means that we can constrain
separations to the order of parsecs, assuming that the precession
  is due to geodetic precession as discussed in Section 1. This is highlighted by the real-world example of Cygnus A, which shows a probable separation distance of \textless 2~pc, with peak in the posterior probability corresponding to a separation of \textless 0.3~pc using the equation mentioned above.

The ability to constrain angular separation and detect binary systems
from jet curvature, using only 20 jet knots, opens up new
possibilities for detecting supermassive black hole binaries using
observational data. We have shown that for jets observed on the arcminute scale, precession periods of around 1 Myr produce better
fits even when inclination angle or cone opening angle do not vary
(Fig.~\ref{fig:truth_all}, all panels). This should not be seen as
surprising, since shorter precession periods will produce a more complex jet morphology, and longer periods will result in straighter jets which are harder to find constraints for. 

Cygnus~A is an excellent example to discuss possible effects of
hydrodynamics. The southwards bend east of 19h59m24s has been
suggested to be likely due to hydrodynamics in a comparison of 3D
hydrodynamic precessing jet simulations by \cite{cox91}.
Interestingly, the ballistic jet models generally do not follow this
bend (compare Figs.~\ref{fig:FullJet} and \ref{fig:StraightJet}). This
supports the hydrodynamic interpretation, but does not rule out the
possibility that the precession is more complex in nature than the
simple conical precession assumed in the models of Section
\ref{subsec:model}.

In magnetised jets, current driven instabilities can dislocate the jet and affect the position of the hotspot \citep{oneill12}. Even when the magnetic field is not dynamically important, the vortex shedding at the jet head introduces a complex feedback loop, impacting the jet directly via the ram pressure of the backflow and via locally increased pressure due to shocks in the jet termination region \citep{lind89}). 3D magneto-hydrodynamical simulations showed that this dynamics leads to some random dislocation of the jet termination regions \citep{mignone10, english16} but the effects were found to also depend on the grid resolution in the simulation \citep{krause01}.

An important characteristic of the results on Cygnus A is that the
fitting is driven by the outer data points located at the primary
hotspots; without this, the model cannot constrain precession periods
since the resulting jet is too straight to provide constraints on the
parameters. Given the impact of the terminal hotspots on the jet
model, it is important to understand whether the position of the
hotspots is driven more by precession, or by hydrodynamics. This will
be explored in future work.

We used a broad flat prior on jet speed (varying $\beta$ from 0.1 to
0.99c), but found that subrelativistic speeds were disfavoured,
contrary to the results obtained by \cite{steenbrugge08}. Their model
requires the southwards bend in the western jet (which we argued above
to be primarily due to hydrodynamic effects) to be explained by the
precession model, while they disregard the eastern hotspot for their fit.
Jet speeds $\beta>0.4$, as found in our analysis, also agree better with VLBI constraints for Cygnus~A \citep[e.g.,][]{boccardi16} and general beaming constraints for jets in radio galaxies \citep{mullin09}.

Although in Cygnus~A we have well constrained jet positions, we have
found that removing down to $\sim 30$ data points makes very little
difference to precession period constraints, and that constraints can
still be found for as few as $\sim 20$ points provided that they adequately
  represent the curvature: as we saw in the case of Cygnus A, removing
  even a few points in strongly curved regions of the jet can have a
  negative effect on the precession period constraints.
  This suggests that the method can be applied to more remote objects with less well-constrained jet positions in future, including sources with less available data compared to Cygnus~A. Given the results of Fig.~\ref{fig:truth_all}, it is worth noting that some remote galaxies with favourable characteristics may still produce a fit at even less than $\sim 20$ points, assuming those are distributed over both jet and counterjet.

Finally we note that the models of Section \ref{subsec:model} work on
the assumption of a constant jet speed. This rules out their direct
application to FRI sources where it is known that the jets
decelerate over large scales \citep{Laing+Bridle14}. The model could
in principle be applied to such sources if a prescription for jet
deceleration were included. Another interesting future improvement might be the inclusion of brightness variations as a consequence of examining the influence of jet speed on apparent surface brightness via Doppler boosting, although this may be challenging in the case of FRII jet sources which typically show non-uniform surface brightness.

\section{Summary \& Conclusions}
The key results for this paper are as follows:

\begin{itemize}
\item We have developed and tested a MCMC model for fitting precession period to images of radio galaxy jets.
\item We have shown that it is possible to find good constraints for precession period in simulated data, in suitable conditions, for varying numbers of data points.
\item The appearance of a counterjet helps stabilise the model and produce better constraints on precession period, even using the same number of points.
\item We applied this to real-world data from Cygnus~A and identified a range of plausible precession parameters. Interpreted in the framework of binary supermassive black hole systems and if precession is caused by the geodetic effect, the binary separation was found to be \textless~0.3 pc.
\item One of the biggest influences on the success of the MCMC code comes from the location of the terminal hotspots. Since this is the case, it is crucial to understand the influence of hydrodynamics in the lobes on the path of the jet.
\end{itemize}

\section*{Acknowledgements}

We thank the anonymous referee for their very useful report that greatly helped to improve the manuscript. MAH acknowledges a studentship from STFC [ST/R504786/1] and MJH acknowledges support from STFC [ST/R000905/1]

\bibliographystyle{mnras}
\bibliography{thesis3}

\bsp
\label{lastpage}
\end{document}